\documentclass[a4paper,fleqn,usenatbib]{mnras}

\usepackage[T1]{fontenc}
\usepackage{ae,aecompl}

\usepackage{graphicx}	
\usepackage{amsmath}	
\usepackage{amssymb}	
\usepackage{multirow}

\newcommand{\nustar}{\textit{NuSTAR}}

\newcommand{\xmm}{\textit{XMM-Newton}}

\newcommand{\nvii}{\ion{N}{VII}}
\newcommand{\ovii}{\ion{O}{VII}}
\newcommand{\oviii}{\ion{O}{VIII}}
\newcommand{\neix}{\ion{Ne}{IX}}

\newcommand{\sixiv}{\ion{Si}{XIV}}

\newcommand{\vsgr}{V4641~Sgr}

\newcommand{\eighteen}{Swift~J1858.6--0814}

\title[Soft X-ray emission lines in Swift J1858.6--0814]{Soft X-ray emission lines in the X-ray binary Swift~J1858.6--0814 observed with \xmm-RGS: disc atmosphere or wind?}

\author[D. J. K. Buisson et al.]{D. J. K. Buisson$^{1}$,\thanks{Email: d.j.k.buisson@soton.ac.uk}
  D. Altamirano$^{1}$,
  M. D{\'i}az Trigo$^{2}$,
  M. Mendez$^{3}$,\newauthor
  M. Armas Padilla$^{4,5}$,
  N. Castro Segura$^{1}$,
  N. D. Degenaar$^{6}$,
  J. van den Eijnden$^{6}$,\newauthor
  F. A. Fogantini$^{7,8}$,
  P. Gandhi$^{1}$,
  C. Knigge$^{1}$,
  T. Mu\~noz-Darias$^{4,5}$, \newauthor
  M. \"{O}zbey Arabac\i$^{1,9}$ and
  F. M. Vincentelli$^{1}$\\
    $^{1}$Department of Physics and Astronomy, University of Southampton, Highfield, Southampton, SO17 1BJ\\
  $^{2}$ESO, Karl-Schwarzschild-Strasse 2, D-85748 Garching bei M\"unchen, Germany\\
  $^{3}$Kapteyn Astronomical Institute, University of Groningen, PO Box 800, NL-9700 AV Groningen, the Netherlands\\
  $^{4}$Instituto de Astrof\'isica de Canarias, 38205 La Laguna, Tenerife, Spain\\
  $^{5}$Departamento de Astrof\'\i{}sica, Universidad de La Laguna, E-38206 La Laguna, Tenerife, Spain\\
  $^{6}$Anton Pannekoek Institute for Astronomy, University of Amsterdam, Science Park 904, 1098 XH, Amsterdam, the Netherlands\\
  $^{7}$Facultad de Ciencias Astron\'omicas y Geof\'{\i}sicas, Universidad Nacional de La Plata, Paseo del Bosque s/n, 1900 La Plata, Argentina\\
  $^{8}$Instituto Argentino de Radioastronom\'{\i}a (CCT-La Plata, CONICET; CICPBA), C.C. No. 5, 1894 Villa Elisa, Argentina\\
  $^{9}$Department of Astronomy \& Astrophysics, Atat\"{u}rk University, Erzurum, Turkey\\
}

\date{Accepted 2020 July 25. Received 2020 July 13; in original form 2020 Feb 18}

\pubyear{2020}

\begin{document}
\label{firstpage}
\pagerange{\pageref{firstpage}--\pageref{lastpage}}
\maketitle

\begin{abstract}
  We find soft X-ray emission lines from the X-ray binary \eighteen\ in data from \xmm-RGS: \ion{N}{VII}, \ion{O}{VII} and \ion{O}{VIII}, as well as notable residuals short of a detection at \ion{Ne}{IX} and other higher ionisation transitions.
  These could be associated with the disc atmosphere, as in accretion disc corona sources, or with a wind, as has been detected in \eighteen\ in emission lines at optical wavelengths. Indeed, the \ion{N}{VII} line is redshifted, consistent with being the emitting component of a P-Cygni profile.
We find that the emitting plasma has an ionisation parameter $\log(\xi)=1.35\pm0.2$ and a density $n>1.5\times10^{11}$\,cm$^{-3}$.
  From this, we infer that the emitting plasma must be within $10^{13}$\,cm of the ionising source, $\sim5\times10^{7}r_{\rm g}$ for a $1.4M_{\odot}$ neutron star, and from the line width that it is at least $10^4r_{\rm g}$ away ($2\times10^{9}(M/1.4M_{\odot})$\,cm).
  We compare this with known classes of emission line regions in other X-ray binaries and active galactic nuclei.
\end{abstract}
\begin{keywords}
  accretion, accretion discs -- X-rays: binaries -- black hole physics -- stars: neutron
\end{keywords}

\begin{figure*}
\includegraphics[width=\textwidth]{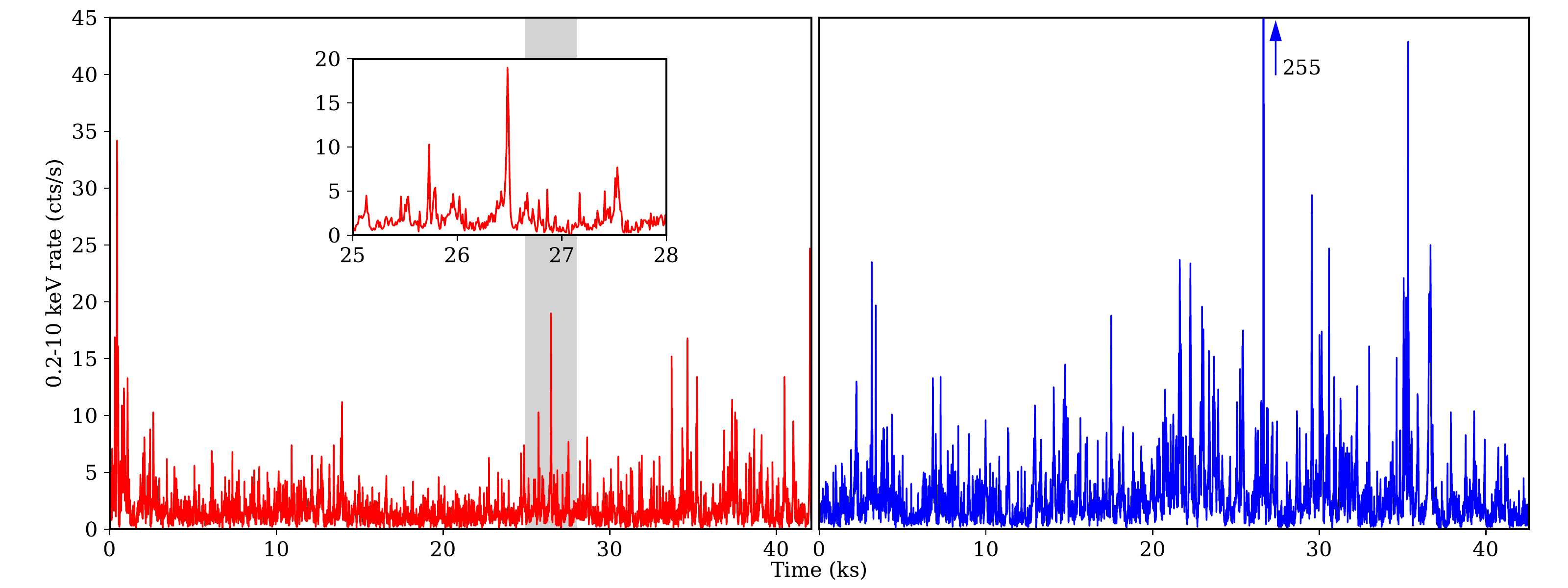}
\caption{\xmm-pn lightcurve of \eighteen. Observations 0831791201 and 0831791701 are shown in red and blue respectively; they are separated by around 20 days. The inset shows a zoom of the region marked in grey. The flare marked with the arrow reaches 255\,cts/s. The frequent, rapid flaring displayed in these observations is typical of \eighteen.}
\label{fig:lightcurve}
\end{figure*}
\begin{figure*}
\includegraphics[width=\textwidth]{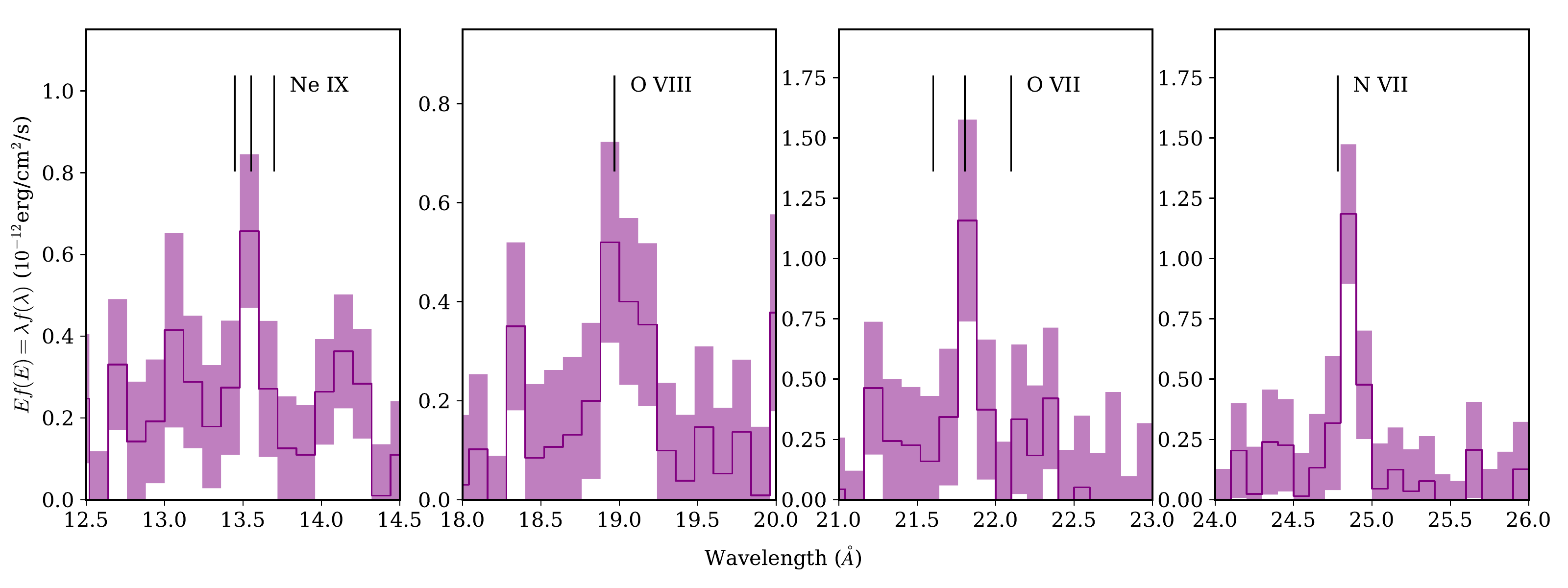}
\caption{Continuum subtracted plots of the mean spectrum of both observations unfolded to a smooth model. Each panel shows a region of the spectrum containing a likely line; the most probable transition, along with its rest wavelength, is indicated.}
\label{fig:lines}
\end{figure*}

\section{Introduction} 

A variety of emission and absorption lines are frequently observed in X-ray binaries; these provide constraints on various aspects of the system.
Several physical components have the necessary conditions to produce these lines: particularly the plasma above the accretion disc.
This may be the optically thin upper layers of the accretion flow, as seen in X-ray emission lines from Accretion Disc Corona (ADC) sources \citep[e.g.][]{cottam01,kallman03}, in which the surface layers of the disc evaporate and become extended \citep{white82,miller00,jimenezgarate02}.
Alternatively, it may be from a disc wind moving away from the central source \citep[e.g.][]{ueda98}. Such winds are an important part of many accreting systems: their power and mass flow rates can be comparable to the luminous power and mass accretion rate \citep{neilsen11,ponti12,casares19} and they can have a significant impact on their environment \citep[e.g.][]{fender16}. However, many aspects of the origin and geometry of winds are still uncertain.
Differentiating when material is ejected in a wind and when it remains bound to the system can show the conditions under which winds occur and hence help to determine how they are driven.

There are several means of driving X-ray binary winds: magnetocentrifugal acceleration of gas \citep[e.g.][]{blandford82,miller06nat}; radiation pressure \citep[e.g.][]{icke80,proga02}; and thermal expansion of the atmosphere \citep[e.g.][]{begelman83,woods96}. To distinguish between these requires observational constraints on plasma properties which differ between the possibilities.

In the X-ray band, winds are most commonly inferred from blue-shifted absorption features \citep[e.g.][]{diaztrigo13}; these include highly ionised species (e.g. \ion{Fe}{XXV}/\ion{Fe}{XXVI}).
Deep observations show that X-ray absorbing winds can consist of several distinct components, with higher ionisation typically occurring at higher velocity \citep[e.g.][]{miller16}.
Sometimes, these features also have evidence for corresponding redshifted emission  \citep[e.g.][]{schulz02,king15,miller15}, giving the full line a P-Cygni profile.

Winds are often also detected as asymmetric emission features from recombination lines in the optical, which may have absorption in their blue wings, \citep{munozdarias16,munozdarias18}, again forming a P-Cygni profile.
The red emission/blue absorption `P-Cygni' profile is expected from outflows such as disc winds \citep[e.g.][]{dorodnitsyn09,dorodnitsyn10,puebla11}
as only material in front of the source, moving towards the observer, absorbs the primary continuum while material moving away from the observer re-emits light in all directions.

Whether the winds detected in the optical and X-ray wavebands are manifestations of the same process or separate phenomena, possibly with different launching mechanisms, is not yet known.
Observing wind properties of the same object in multiple wavebands is key to resolving this.

\subsection{\eighteen}

\eighteen\ is a recently discovered \citep{krimm18} X-ray binary which shows P-Cygni line profiles in its optical spectra \citep{munoz19,munoz20}.
Its outburst has been long-lasting, being first detected on 25$^{\rm th}$ October 2018 \citep{krimm18} and ongoing at the time of writing \citep{hare19,rajwade19,buisson20atel1}, which has provided the opportunity for extensive observations with many instruments across a wide range of bandpasses. 
Recent detections of Type I X-ray bursts show that \eighteen\ has a neutron star accretor (Buisson et al. submitted); these bursts exhibit photospheric radius expansion, allowing a distance estimate of $\sim15$\,kpc; and periodic eclipses have been found, showing that \eighteen\ is viewed at high inclination \citep{buisson20atel2}.
Before 2020, a common property of essentially all observations was atypically large variability.

Radio emission showed variability by up to a factor of $7-8$ within 20 minutes and is consistent with a compact jet; the variability is likely due to mass accretion rate fluctuations rather than discrete ejecta \citep{bright18,vandeneijnden20}.
Optical variability included changes by factors of several within minutes \citep{vasilopoulos18} and fast flares, which are stronger at longer wavelengths \citep{paice18}. The wind absorption also changed in strength and terminal velocity, with changes not clearly related to the continuum variability \citep{munoz20}. Such changes also occur in other systems, \citep[e.g.][]{prinja00,kafka04}.
 
The rapid X-ray variability in \eighteen\ was extremely strong, occasionally changing by a factor of more than two hundred within a few seconds (Fogantini et al. in prep.), which contrasted with the relatively stable long-term average flux.
The X-ray spectra of \eighteen\ are also unusual in their hardness ($\Gamma\sim1.5$) and show strong, variable obscuration \citep{reynolds18,hare20}.
This is reminiscent of previous outbursts of the unusual X-ray binaries \vsgr\ \citep{wijnands00,revnivtsev02} and V404~Cyg \citep{gandhi16,walton17,motta17abs}.

In this paper, we present high resolution soft X-ray spectra of \eighteen\ from the \xmm\ Reflection Grating Spectrometer (RGS) . We describe the available observations and their reduction in Section~\ref{sec:odr}; give results in Section \ref{sec:res}; and consider their implications in Section~\ref{sec:dis}.

\section{Observations and data reduction}
\label{sec:odr}
We analyse the two \xmm\ \citep{jansen01} observations of \eighteen, focussing on the RGS data. These have OBSIDs 0831791201 and 0831791701, taken on 2018 March 23 and April 13 respectively.
Hereafter, we refer to these as observations 201 and 701.
We reduced the RGS data with the \xmm\ SAS, version 17.0.0.
Using \textsc{rgsproc}, we extracted source spectra from regions including 95\% of the PSF and the background from outside of 99\%.
We do this for both RGS 1 and 2 and spectral orders 1 and 2.
This produced spectra with exposures of 58 and 45\,ks respectively.

To retain the maximum information in the spectrum, we use the unbinned data, discarding channels labelled by \textsc{rgsproc} as bad.
When fitting, we leave the different cameras, spectral orders and observations as separate data sets, fitting jointly where relevant.
The low counts in each channel necessitates using the statistic of \citet{cash79}, which treats the Poisson data accurately. We perform fits in \textsc{isis} version 1.6.2-43 \citep{houck00}.

We also extract a $0.2-10$\,keV light curve from the pn data for illustrative purposes, using \textsc{evselect} and \textsc{epiclccorr}. We include events with PATTERN<=4, using columns $33-41$ for the source and $55-63$ as background (the observation was taken in timing mode).

\begin{figure*}
\includegraphics[width=\textwidth]{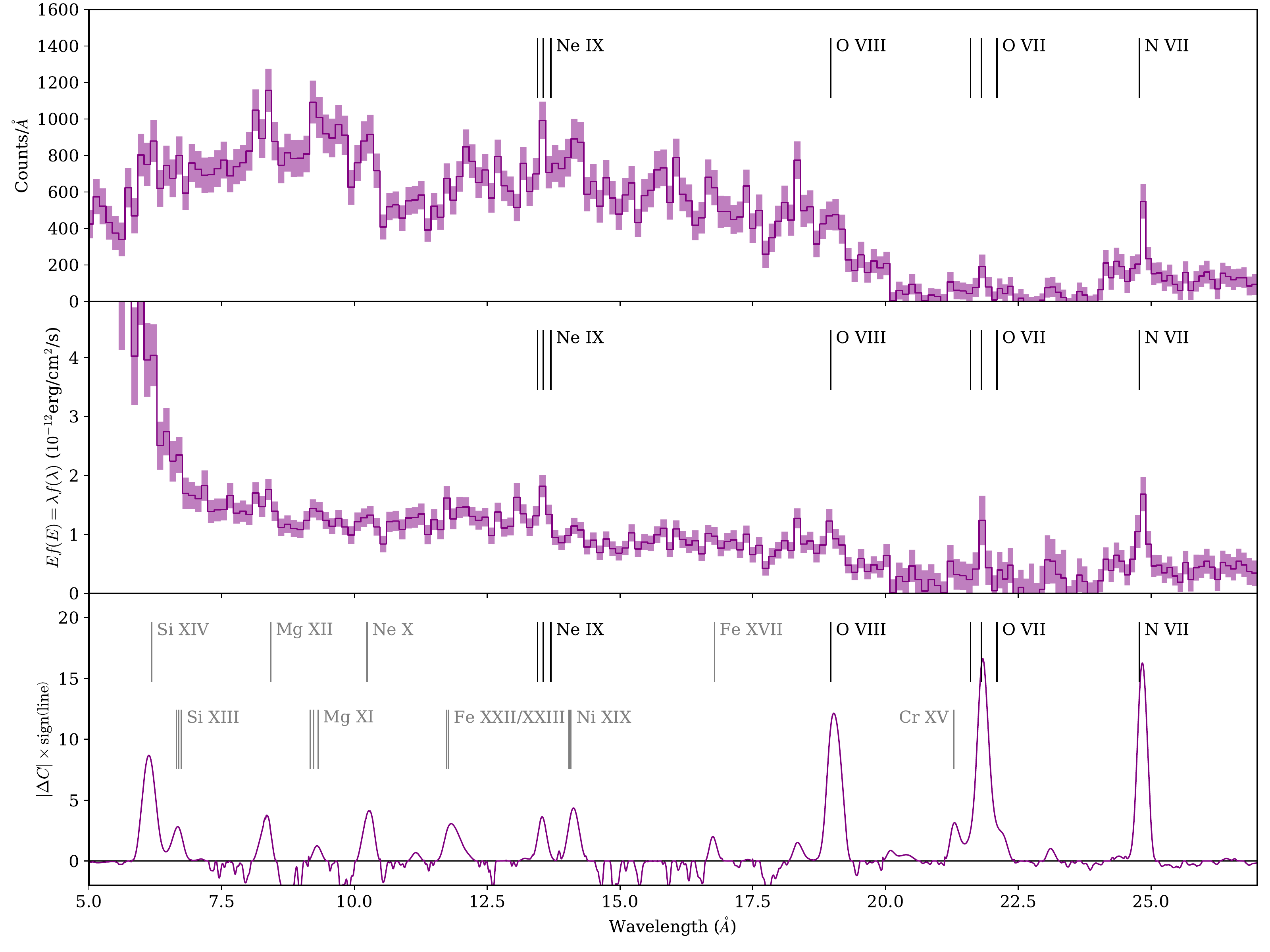}
\caption{Combined RGS spectrum of \eighteen\ (rebinned for plotting). Top: Background subtracted spectrum of both observations combined, shown as detected counts summed over all detectors and orders. Second row: unfolded to a $\Gamma=2$ power law to remove the effect of the detector effective area.
Bottom row: improvement in fit statistic from adding a narrow Gaussian to a continuum model. The magnitude gives $\Delta C$ and the sign shows the nature of the additional line: positive represents additional emission; negative represents absorption.
\ion{O}{VII} and \ion{N}{VII} are clearly detected; given these, \ion{O}{VIII} is detected at the expected wavelength. Plausible transitions (grey) exist for all features with $\Delta C>4$.}
\label{fig:unfold}
\end{figure*}

\section{Results}
\label{sec:res}

We begin by showing the light curve of each observation (Figure~\ref{fig:lightcurve}). This shows strong, rapid flaring as occurs throughout the outburst of \eighteen. This flaring dominates the variability while the long-term flux is more stable. Therefore, we initially produce a mean spectrum for the whole campaign (the effect of variability on the spectrum is considered in Section~\ref{sec:line_sig}).
The most prominent features in the RGS spectrum are several apparent emission lines (Figure~\ref{fig:lines}); the significance of these lines and the conclusions which can be inferred from them are the focus of this paper.
To give a visual overview of the spectral shape, we also show the mean RGS spectrum of both observations combined summed across both detectors and orders, as raw counts and unfolded to a constant model (in $\nu F\nu$, i.e. a $\Gamma=2$ power law), in the upper and middle panels of Figure~\ref{fig:unfold}.

\subsection{Identification of emission lines}
\label{sec:lines}

\begin{figure}
\includegraphics[width=\columnwidth]{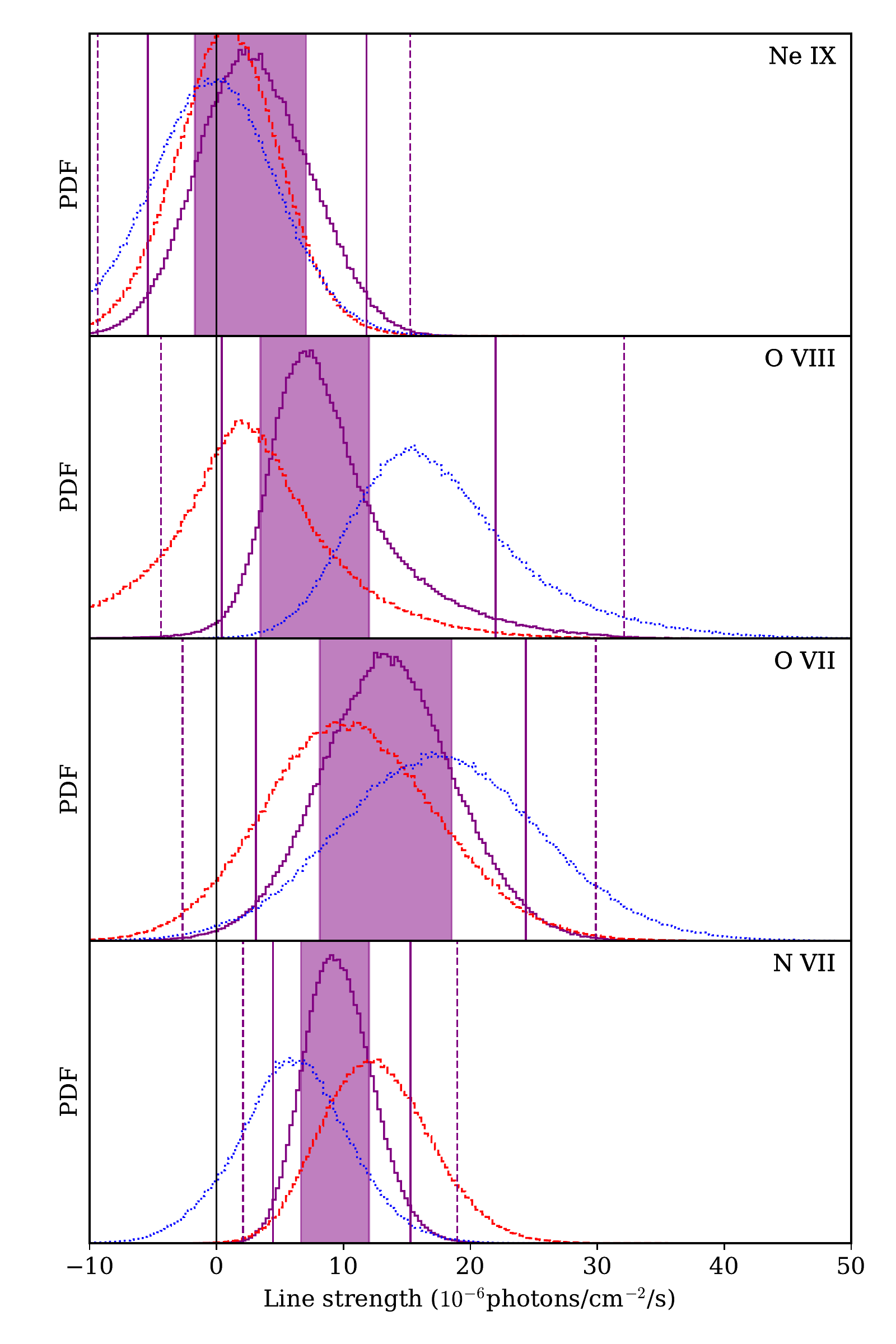}
\caption{Posterior distributions of emission line strengths. For He-like triplets, the sum of all components is shown. Obs~2 is shown in red dashed; Obs~7 in blue dotted; and the combined observation in purple. Shaded regions, solid and dashed lines denote 68, 95 and 99.7\% HPD credible regions respectively.}
\label{fig:line_strength}
\end{figure}

To consider the significance of the emission lines in an unbiased way, we begin by performing a line scan, adding an unresolved Gaussian of fixed wavelength to a smooth continuum and recording the fit improvement at each wavelength. We use a continuum of a blackbody, since this was found to be the dominant spectral component at these energies by \citet{hare20}. We also tried including a powerlaw (which is dominant at higher energies) or using an absorbed disc-blackbody, but this turns out to make negligible difference.

The resulting fit improvements are shown in the bottom panel of Figure~\ref{fig:unfold}.
This shows clearly detected lines at around the rest wavelengths of \ovii\ and \nvii.
Since these are known wavelengths of two of the most common lines, one need not consider the look-elsewhere effect for the full wavelength range tested.
There is also a weaker feature at \oviii;  the fit improvement would correspond to approximately $3\sigma$ if treating the change in statistic as a simple likelihood ratio but since this is not always reliable, \citep[e.g.][]{protassov02} more careful treatment of the significance is required.
Furthermore, there are many features with a change of fit statistic, $\Delta C$, between 4 and 9. These are not robust detections as the number of wavelengths tested (several hundred resolution elements) would be expected to produce several such features by chance. However, they will still play a role in determining the best-fitting plasma models so we note the ion of the common transition which fits each such feature most closely in Figure~\ref{fig:unfold}.

\subsection{Significance and properties of lines}
\label{sec:line_sig}

We now perform a more detailed analysis of the most likely lines. We consider the clearly detected lines ($\Delta C>9$) -- \ion{O}{viii}, \ion{O}{vii} and \ion{N}{vii} -- along with the strongest tentative line, \ion{Ne}{IX} (apart from \sixiv, which is at the edge of the bandpass).

For each line, we use local models of a smooth continuum along with a narrow line.
We fit a power law to the data within the region covering  $\pm2$\,\AA\ of the line, with a Gaussian allowed to be broadened and have a small redshift from the rest wavelength of the line.
For the He-like ions (\neix\ and \ovii), we use a triplet of three Gaussians with a common width and redshift, but independent strengths.
We generate posterior distributions for line parameters with Markov Chain Monte-Carlo (MCMC) chains. For each line, we produce a MCMC chain for each observation separately and for both observations combined.
We generate chains with 100 walkers per free parameter and 5000 steps, using the implementation in ISIS of the implementation of \citet{foremanmackey13} of the method of \citet{goodman10}. We use flat (uninformative) priors on each parameter.
We allow the line strength to be positive or negative such that the resulting posterior distribution gives the probability of an emission feature.
The resulting parameters are shown in Table~\ref{tab:lines}; we do not show \neix\ since the inferred line strength includes 0, so the width and redshift are not meaningful.
We find that the probabilities of each line showing emission match the order of the $\Delta C$ values: \neix\ has only a 75\% chance of being an emission line; the \nvii\ line is the most likely at 99.97\%.

\begin{table*}
\caption{Properties of emission lines in the combined RGS data. Errors are 90\% credible intervals.}
\label{tab:lines}

\begin{tabular}{lccc}
\hline
Ion &  O VIII & O VII & NVII \\
Transition & 1s$^1-$2p$^1$ & 1s$^2-$1s$^1$2p$^1$ & 1s$^1-$2p$^1$     \\
\hline
$P($Line strength $>0)$ &  0.989 & 0.994 & 0.9997 \\
Line strength/$10^{-6}$\,photons\,cm$^{-2}$\,s$^{-1}$ &   $8_{-7}^{+10}$  &  $13_{-8}^{+9}$  &  $9_{-4}^{+5}$ \\
\hline
Redshift/$10^{-3}$ &   $1\pm9$  &  $1.5_{-4.2}^{+5.6}$  &  $2.2_{-2.0}^{+2.2}$  \\
Velocity/km\,s$^{-1}$ &  $400_{-2700}^{+2900}$  &  $450_{-1250}^{+1700}$  &    $650_{-600}^{+680}$ \\
\hline
Width/\AA &   $<0.27$  & $<0.11$  &  $<0.13$ \\
Width/km\,s$^{-1}$ &  $<4300$  &  $<1600$  &  $<1600$ \\
\hline
\end{tabular}

\end{table*}

We also extract posteriors on the width and redshift of the lines. We find that in most cases, the values are consistent with being at rest.
However, the \nvii\ line shows evidence for being redshifted ($P(z>0)=0.94$) and if the lines are assumed to have the same redshift, the combined probability reinforces this to $P(z>0)=0.96$.
This may have a contribution from instrumental systematic errors or the source radial velocity but if intrinsic to the source could be because the full line profile is a P-Cygni profile from a wind, with the absorbed blue side of the line shifting the line centroid redward.

We also test for variability of the detected lines by comparing the posteriors between the two observations.
For each species, the 90\% credible intervals from the two observations overlap, so the line strengths are consistent with remaining constant between the observations. Similarly, for no line is there a significant probability that one particular observation is brighter than the other (the probabilities that Obs~201 has a brighter line than Obs~701 being: 0.43 for \neix; 0.08 for \oviii; 0.17 for \ovii; and 0.86 for \nvii).
We also search for variability of the lines within an observation, in particular with the continuum flux. We do this by splitting the RGS exposure based on the $0.7-10$\,keV flux from the pn data.
Even when split into only two flux states, this does not place interesting constraints on the variability in terms of either line flux or equivalent width.

\subsection{Gas density from the O\,VII triplet}

To derive physical properties of the plasma, we can either consider the most important transitions individually or use models of the full spectrum.
Full models maximise the spectral information used and reduce the effect of  possible contamination of an individual feature but require more assumptions (for example relative abundances) about the plasma.
We begin by considering parameters which may be constrained using the most strongly detected transitions 
and model the full spectrum in Section~\ref{sec:phot}.

The \ion{O}{VII} feature at around 22\,\AA\ is actually a triplet, with components at 21.60, 21.80 and 22.10\,\AA. The relative strength of these components is sensitive to the ionisation mechanism and density of the gas \citep{porquet01}.
The resonance (r, $w$, 22.1\,\AA) line is strong in collisional-dominated (thermally ionised) plasmas and weak in photo-ionised plasmas.
The relative strength of the forbidden (f, $z$, 21.6\,\AA) and intercombination (i, $x+y$, 21.8\,\AA) lines depends on the density, with forbidden being strong at low density and intercombination and high density (the relevant density range is $n_{\rm e}=10^{10}-10^{12}$\,cm$^{-2}$).
Here, the intercombination line is strongest, implying high-density photoionised plasma (since \oviii\ and \nvii\ are detected with at most a small wavelength shift, it is unlikely that the apparent intercombination line is a blueshifted forbidden line).

\begin{figure}
\includegraphics[width=\columnwidth]{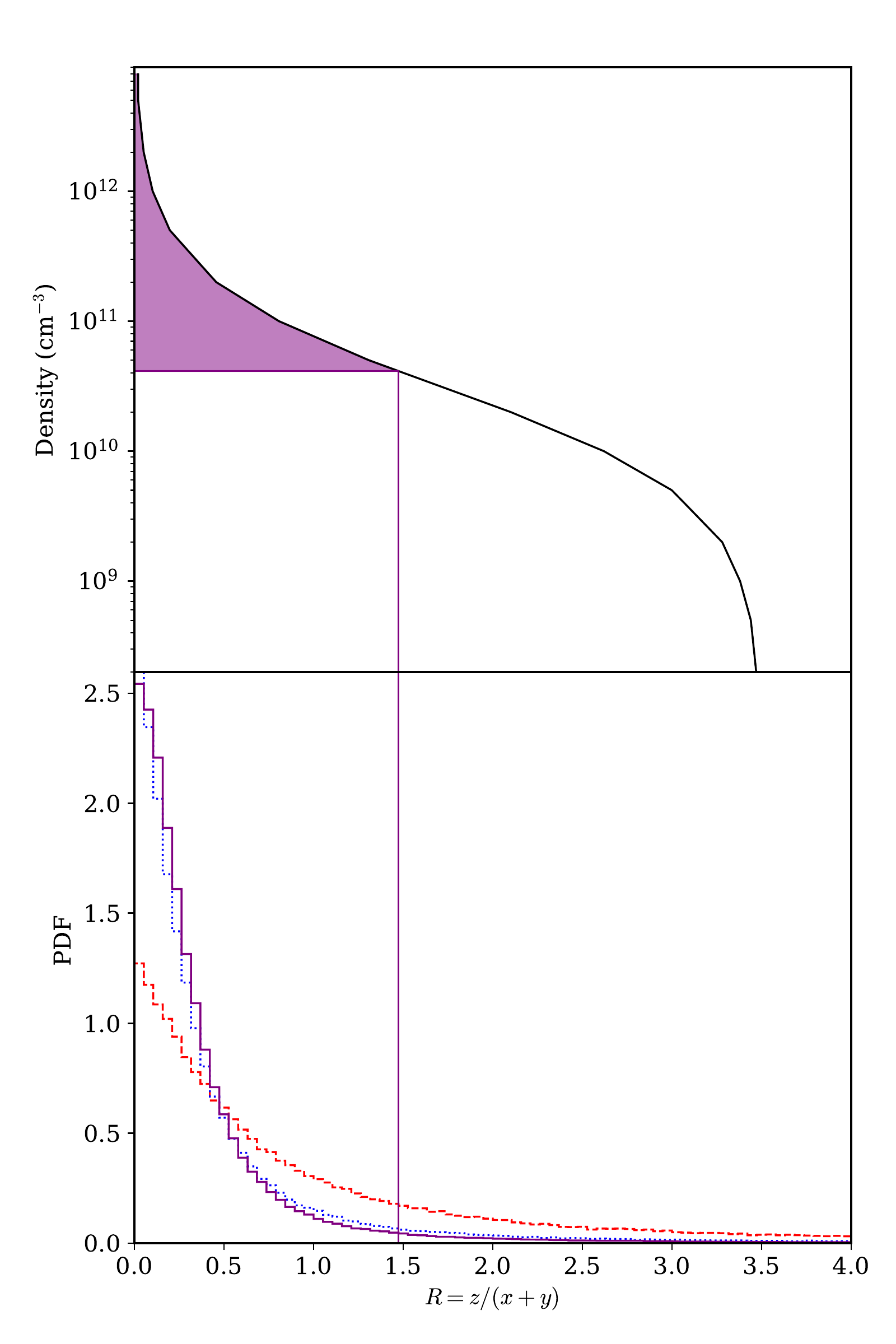}
\caption{Constraint on density from the \ion{O}{VII} triplet. Bottom panel: probability density of the forbidden ($z$) to intercombination ($x+y$) ratio; the vertical line marks the 95\% upper limit.  (Obs~2 is shown in red dashed; Obs~7 in blue dotted; and the combined observation in purple). Top panel: relation between line ratio and density, showing the corresponding lower limit to the density.}
\label{fig:density}
\end{figure}

We use the theoretical line ratios calculated by \citet{porquet01} to quantify this, comparing with the measured line ratio from the MCMC chains calculated in Section~\ref{sec:lines}. Figure~\ref{fig:density} shows the constraint on density, from the ratio $R=z/\left(x+y\right)$.
Since the forbidden line is not detected, the ratio $R$ is only constrained as an upper limit, which corresponds to a  lower limit on the density.
This shows that $n_{\rm e}>4.1\times10^{10}$\,cm$^{-3}$ at 95\% credibility.

We also apply this method to the \ion{Ne}{IX} line, but the line detections are not strong enough to provide a density constraint.

\subsection{Ionisation from O VIII/O VII ratio}
\label{sec:ion}

\begin{figure}
\includegraphics[width=\columnwidth]{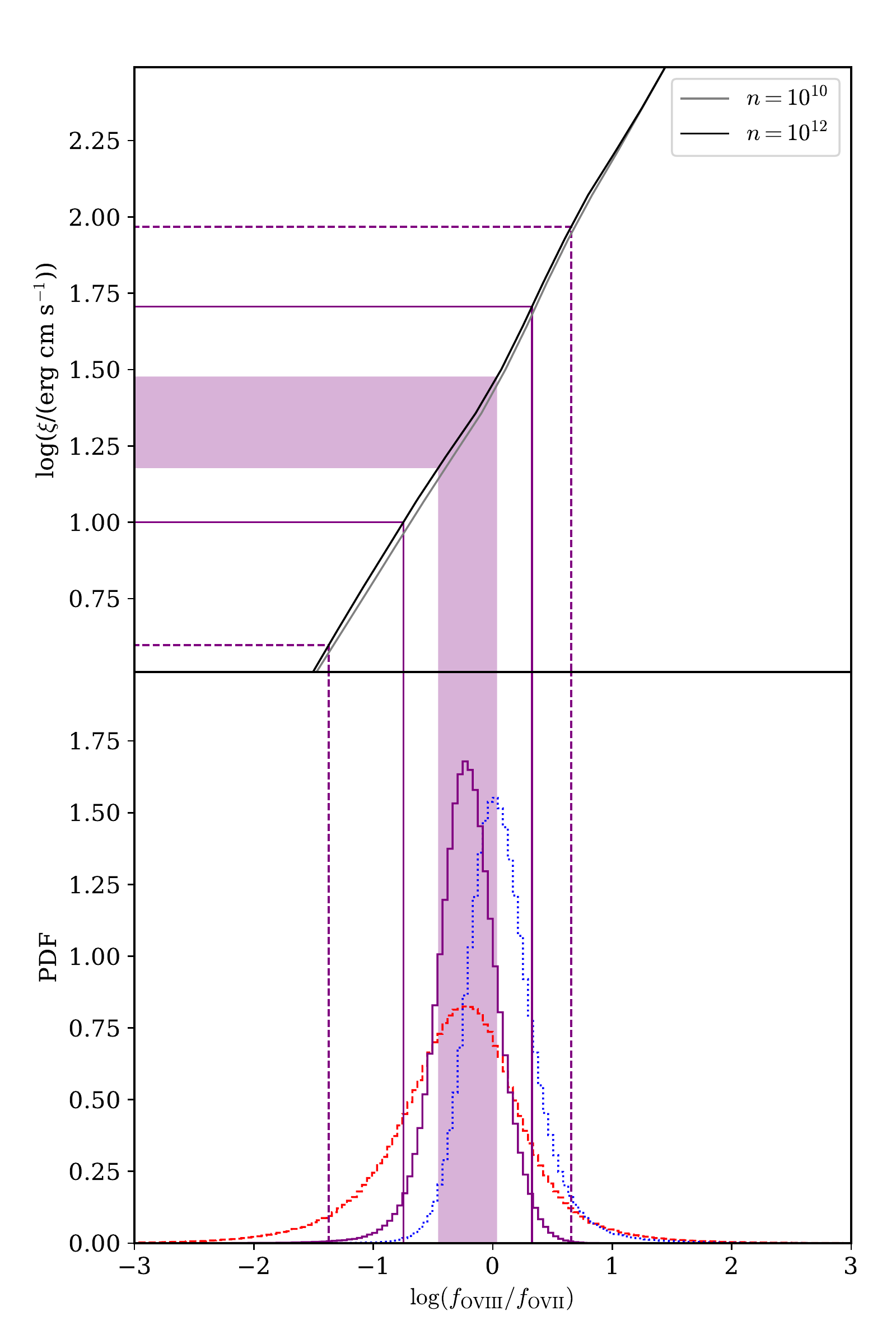}
\caption{Constraint on the ionisation parameter from the \ion{O}{VIII} to \ion{O}{VII} ratio. Bottom panel: probability density of the line flux ratio (Obs~2 is shown in red dashed; Obs~7 in blue dotted; and the combined observation in purple); the shaded regions and vertical lines mark 1$\sigma$ (68\%) and 2$\sigma$ (90\%) credible intervals respectively for the combined observation. Top panel: relation between line ratio and ionisation parameter, showing the corresponding constraints on the ionisation. The black and grey lines show plasma densities of $n=10^{10}$ and $n=10^{12}$\,cm$^{-3}$ respectively; this has only a very small effect.}
\label{fig:ionisation}
\end{figure}

The degree of ionisation of the plasma is reflected in the relative strengths of lines from different ionisation states of the same element. Here, we can use the relative flux in the \ion{O}{VII} and \ion{O}{VIII} lines.

We calculate the relative strengths of the \ion{O}{VII} and \ion{O}{VIII} lines at different ionisation parameters using the \textsc{xstar} package, version 2.5 \citep{kallman01}.
We calculate spectra over the range $\xi=3-300$\,erg\,cm\,s$^{-1}$ in 20 logarithmically spaced steps and extract the flux for each line from the line luminosity file.
We use an illuminating continuum based on fits to the \nustar\ spectrum \citep{hare20} of observations in the same state as analysed here: a cut-off powerlaw with $\Gamma=1.5$ and $E_{\rm Cut}=30$\,keV. We use Solar abundances and a density $n=10^{12}$\,cm$^{-3}$.

We then compare these predicted ratios with the ratio measured in the MCMC chains produced in Section~\ref{sec:lines}. The distribution for each observation is shown in Figure~\ref{fig:ionisation}. This gives $\log(\xi)=1.2_{-0.6}^{+0.5}$ for observation 201, $\log(\xi)=1.4_{-0.2}^{+0.3}$ for observation 701 and $\log(\xi)=1.3_{-0.2}^{+0.3}$ (90\% HPD) for the combined dataset.

We also test the effects of density by repeating this process for $n=10^{10}$\,cm$^{-3}$ but find that this has a very weak effect (Figure~\ref{fig:ionisation}). While the components of the \ion{O}{VII} triplet change, their total flux remains similar in comparison to \ion{O}{VIII}.

\subsection{Photoionised emission modelling}
\label{sec:phot}

\begin{figure*}
\includegraphics[width=\textwidth]{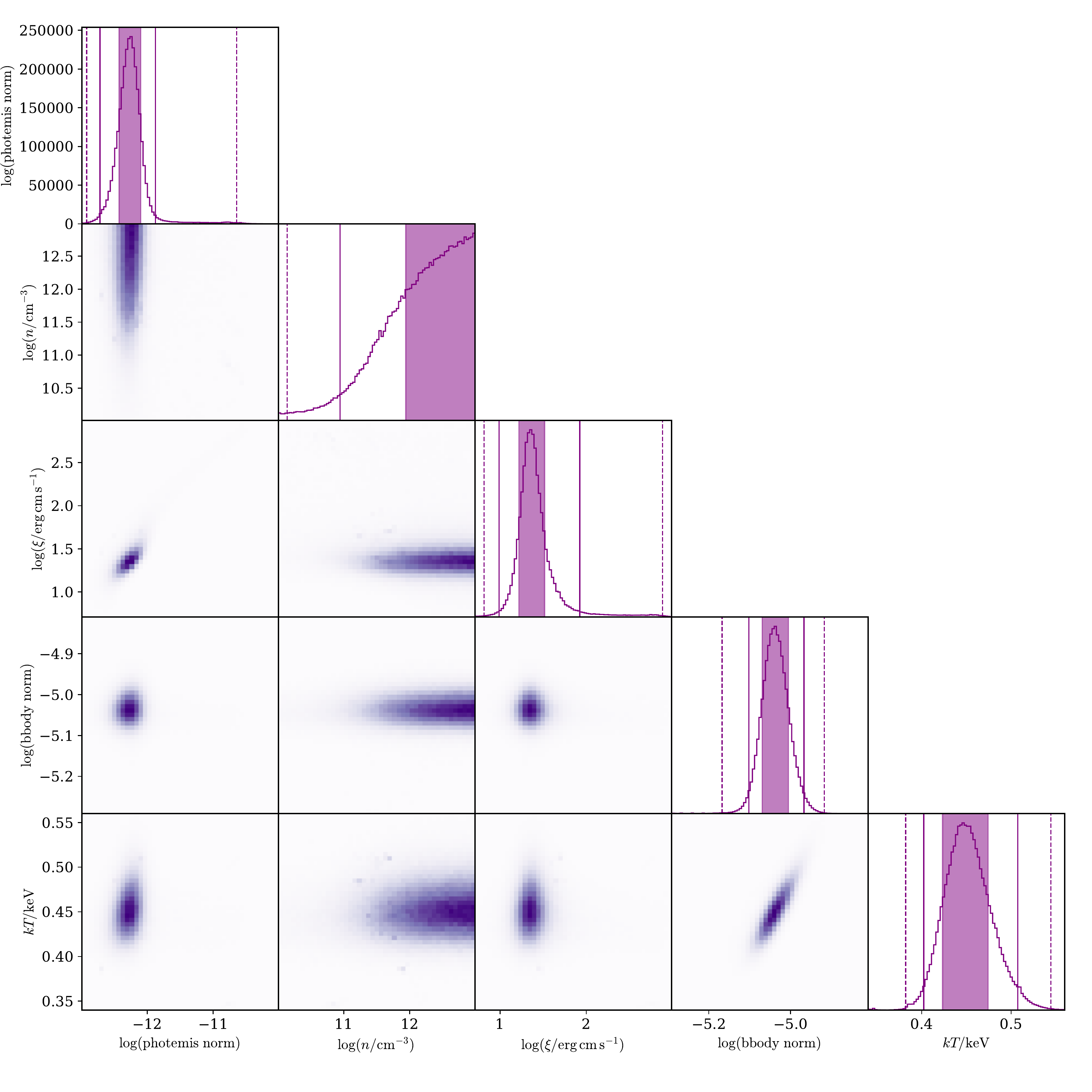}
\caption{Posterior density plots from photoionisation modelling. Normalisations are the \textsc{xspec} values. Shaded regions, solid and dashed lines denote 68, 95 and 99.7\% HPD credible regions respectively.}
\label{fig:phot}
\end{figure*}

\begin{table*}
\caption{Parameters of fits to photoionisation models. Normalisations are the \textsc{xspec} definitions; errors are given at 90\% confidence.
We also show the values derived in previous sections from individual lines, which are consistent with the results from full photoionisation modelling.}
\centering
\begin{tabular}{lcccc}
\hline
Dataset & Obs 201 & Obs 701 & Combined & Combined \\
\hline
Method & \multicolumn{3}{c}{Photoionisation model fitting} & From individual lines\\
\hline
${\rm log}(n/{\rm cm^{-3}})$ &  $>10.6$  &  $>10.5$ & $>11.3$ & $>10.6$ (95\%) \\
${\rm log}(\xi/{\rm erg\,cm\,s^{-1}})$ &  $1.3_{-0.5}^{+1.2}$  &  $1.5_{-0.4}^{+0.5}$  & $1.3_{-0.2}^{+0.4}$ & $1.3_{-0.2}^{+0.3}$ \\
Blackbody $kT$/keV &  $0.43_{-0.05}^{+0.06}$  &  $0.47_{-0.06}^{+0.07}$ & $0.44_{-0.03}^{+0.05}$ & $-$\\
\hline
\end{tabular}
\label{tab:phot}
\end{table*}

As well as constraining the plasma properties from individual lines, we use a photoionisation model to model the full spectrum.
Using \textsc{xstar}, we construct a table model for the emission component with density and ionisation as free parameters. We use 25 logarithmic steps over $n=10^{10}-10^{13}$\,cm$^{-3}$ and $\xi=3-1000$\,erg\,cm\,s$^{-1}$, with the same input continuum and ancillary parameters as in Section~\ref{sec:ion}.

We fit a single instance of the emission component from this grid, along with a \textsc{blackbody+powerlaw} continuum model. However, the powerlaw component is consistent with having zero flux (within 90\% errors) so we do not consider it further.
The fit improvement of the model with emission over pure continuum is $\Delta C=25$.
We also test the addition of a second photoionised emission component but do not find a significant improvement ($\Delta C=1.7$ for 3 degrees of freedom).

Best-fit parameters are given in Table~\ref{tab:phot}. The values of density, $\log(n/{\rm cm^{3}})>11.2$, and ionisation parameter, $\log(\xi/{\rm erg\,cm\,s^{-1}})=1.35\pm0.2$, for the combined dataset are consistent with those measured from individual transitions. 

We also perform a MCMC analysis based on this model. The posterior density for each parameter is shown in Figure~\ref{fig:phot} and the median and HPD credible intervals in Table~\ref{tab:phot}. This shows that there are no strong degeneracies between parameters and the derived values are similar to those found with traditional fitting methods. 

\subsection{Derived properties}

From the quantities we have already measured, we may also place constraints on other important physical properties of the system.
The distance from the source to the emitting plasma can be inferred by combining the luminosity, density and ionisation parameter:
$$R=\sqrt{L/n\xi}$$
We have already measured the ionisation and placed a lower limit on the density. The distance to \eighteen\ has been estimated from the flux during photoshperic radius expansion of Type I X-ray bursts \citep[Buisson et al. submitted]{buisson20atel2} as around 15\,kpc.
This gives a luminosity $L\sim2\times10^{38}$\,erg\,s$^{-1}$ (\nustar\ spectra taken at the times of the \xmm\ spectra show a similar flux to that analysed in \citealt{hare20}).
Combining this with the lower limits to density and ionisation from observation 701 gives $R\lesssim10^{13}(D/15\,{\rm kpc})$\,cm, corresponding to $5\times10^7r_{\rm g}$ for a $1.4M_\odot$ neutron star.

A lower limit to the proximity of the emitting gas can be set by considering the amount of broadening observed in the line.
Taking the line width as comparable to the Keplerian velocity gives $r/r_{\rm g}\sim \left(c/v\right)^2$; since the line may also be broadened by other effects, this estimate is a lower limit to the proximity of the emitting gas. For the most strongly constrained \nvii\ line, $r>4\times10^{4}r_{\rm g}=8\times10^{9}(M/1.4M_{\odot})$\,cm.

\section{Discussion}
\label{sec:dis}

We have found photoionised gas in \eighteen\ with emission lines from \ion{Ne}{IX}, \ovii, \oviii\ and \nvii. From the properties of these lines, we can infer various physical parameters of the gas and its role in the accretion system of \eighteen; it is likely to either be a hot accretion disc atmosphere or a component of the wind which is seen in other wavebands. The various constraints are discussed below.

We find that the density of the emitting plasma is $n_{\rm e}\geq 1.5\times10^{11}$\,cm$^{-3}$.
While the effect of UV radiation on He-like triplet line ratios may bias our density measurement, high densities have also been found in X-ray absorbing winds from around black hole XRBs using \ion{Fe}{XXII} line ratios (11.77\,\AA,11.92\,\AA; see \citealt{mauche03}), e.g. $\sim10^{14}$\,cm$^{-3}$ \citep{miller06nat,miller08}; $10^{16-17}$\,cm$^{-3}$ \citep{miller14}.
Our limit is also consistent with other measurements of emitting plasma in X-ray binaries \citep{cottam01, kallman03,psaradaki18} and is dense compared to many astrophysical objects.
These constraints are however far from a typical binary disc value $\sim10^{20}-10^{22}$\,cm$^{-3}$ \citep{svensson94,tomsick18,jiang19} so the emission is unlikely to be from the photosphere itself.

A similar emission component to that found here is detected by \citet{psaradaki18} in the neutron star LMXB EXO~0748--676: they find a similar ionisation to that found here and use several He-like triplets to infer a plasma density of $>10^{13}$\,cm$^{-3}$.
These lower ionisation components (compared to many other XRBs), may be common but less studied due to Galactic absorption \citep{costantini12,diaztrigo16}.
\citet{psaradaki18} also find a higher ionisation ($\log(\xi/{\rm erg\,cm\,s^{-1}})\sim2.5$) component, which could have an equivalent in \eighteen\ responsible for the residuals matching higher ionisation transitions that we find. A deeper high-resolution soft X-ray spectrum would be required to test this.

The upper limit on the distance ($10^{13}$\,cm$\sim5\times10^{7}r_{\rm g}$) of the emitting plasma from the neutron star is not highly constraining. For example, a typical launching radius seen in absorbing winds is $10^{3}-10^{4}r_{\rm g}$ \citep[e.g.][]{kallman09}. Similarly, where a disc wind is invoked as the material for the broad line region in AGN, it typically exists at around $10^{3-4}r_{\rm g}$ \citep{tremaine14}.
However, these sizes  are typically for absorbing material detected with higher ionisation, which is likely to occur closer to the central source: the emission observed here could very well be located further from it.

We also have a lower limit on the size of the emitting region from the narrowness of the lines: this implies that the majority of the \nvii\ emission comes from $r>4\times10^4r_{\rm g}$. It is possible that the lines are emitted closer to the neutron star with broader profiles but their blue wing is absorbed. However, absorbing the whole of the blue side would only allow for a change by a factor of $\sim2$ in velocity and $\sim4$ in radius.
This distance constraint is at the upper end of AGN BLR radii, so the component we observe may be closer to an analogue of an AGN NLR (in terms of scale in $r_{\rm g}$; AGN NLR densities are much lower, typically $<10^5$\,cm$^{-3}$ \citealt{bennert06}).

As well as the size, we can also consider the shape of the emitting material.
Disc winds in binaries are predicted to include material ejected in all directions \citep[e.g.][]{higginbottom18}; the denser regions producing X-ray signatures are believed to be predominantly equatorial, since they are usually seen in absorption in high inclination sources \citep{boirin05,miller06nat,miller061743, king12,ponti12}.
\eighteen\ shows periodic eclipses, which shows that it is at high inclination \citep[$>70^\circ$,][Buisson et al. in prep.]{buisson20atel2}.
This can explain the emission line properties \citep[c.f.][]{charles19}: if much of the X-ray continuum is blocked by the disc, even a modestly high scattering atmosphere will produce observable emission lines.
In this situation, the unusually strong variability of \eighteen\ can be explained as variable obscuration from surface of a rough disc viewed at very high inclination: as the material in the disc orbits, thicker obscuring regions move into and out of the line of sight. Also, the surface of the disc will include more dense material, which can maintain sufficiently low ionisation to produce optical absorption features. This scenario may be able to explain the observed optical wind \citep{munoz20}.

We also find some evidence for the emission lines being redshifted ($P(z>0)=0.96)$), largely due to the \nvii\ line ($P(z>0)=0.96)$).
This may include contributions from the absolute calibration of the RGS and the radial velocity of the source.
The RGS is calibrated to 6\,m\AA\footnote{https://xmm-tools.cosmos.esa.int/external/xmm\_user\_support/ documentation/uhb/rgswavscal.html}, corresponding to $\pm80$\,km\,s$^{-1}$ at the wavelength of \nvii.
There is no evidence for significant radial velocity in the optical spectra \citep{munoz20} and most X-ray binaries have a radial velocity of $\lesssim100$\,km\,s$^{-1}$ \citep{gandhi19}.
Allowing a conservative $200$\,km\,s$^{-1}$ for these effects still leaves a likely redshift, $P(z>200\,{\rm km\,s^{-1}})=0.89$.
If there is an intrinsic source redshift, it could be due to the blue wing of the line being absorbed to form a P-Cygni profile as is also seen in the optical spectra \citep{munoz20}.
If this is the case, the redward extent of the line can be used to estimate the wind velocity: combining the width of the line with its redshift gives $v\sim2000$\,km\,s$^{-1}$, similar to other wind velocities.
An alternative interpretation is that the line is emitted by gravitationally redshifted or infalling material:
\citet{miller14} find (although in a compromised observation) redshifted absorption components in MAXI~J1305--704 corresponding to the free-fall and gravitational redshift of material at around $500r_{\rm g}$.
Since the lines observed here are seen in emission, an inflow is less likely since there would need to be a large central structure blocking the blueshifted emission from the far side.
Alternatively, the redshifted emission could occur where the accretion stream impacts the disc, as suggested by \citet{psaradaki18}.

To directly confirm the X-ray P-Cygni profile by observing the absorption would require unfeasibly long RGS observations, but could be possible with brighter sources or future telescopes with larger effective areas.

\section*{Acknowledgements}

We thank the referee for useful comments which have improved the clarity of the manuscript. DA and DJKB acknowledge support from the Royal Society.
MAP is funded by the Juan de la Cierva Fellowship IJCI--2016-30867. TMD is funded by the Ram\'on y Cajal Fellowship RYC-2015-18148. MAP and TMD acknowledge support by the Spanish MINECO grant AYA2017-83216-P. 
NCS and CK acknowledge support by the Science and Technology Facilities Council (STFC), and from STFC grant ST/M001326/1.
JvdE and ND are supported by a Vidi grant from the Netherlands Organisation for Scientific research (NWO), awarded to ND.
FF and DA acknowledge support from the Royal Society International Exchanges "The first step for High-Energy Astrophysics relations between Argentina and UK".This work has made use of observations obtained with \xmm, an ESA science mission with instruments and contributions directly funded by ESA Member States and NASA.
This research has made use of ISIS functions (ISISscripts) provided by ECAP/Remeis observatory and MIT (http://www.sternwarte.uni-erlangen.de/isis/).
MOA acknowledges support from the Royal Society through the Newton International Fellowship programme.
FMV acknowledges support from STFC under grant ST/R000638/1

\section*{Data availability}

The data on which this work is based are available from the XMM-Newton Science Archive.

\bibliographystyle{mnras}
\bibliography{swiftj1858}

\begin{thebibliography}{}
\makeatletter
\relax
\def\mn@urlcharsother{\let\do\@makeother \do\$\do\&\do\#\do\^\do\_\do\%\do\~}
\def\mn@doi{\begingroup\mn@urlcharsother \@ifnextchar [ {\mn@doi@}
  {\mn@doi@[]}}
\def\mn@doi@[#1]#2{\def\@tempa{#1}\ifx\@tempa\@empty \href
  {http://dx.doi.org/#2} {doi:#2}\else \href {http://dx.doi.org/#2} {#1}\fi
  \endgroup}
\def\mn@eprint#1#2{\mn@eprint@#1:#2::\@nil}
\def\mn@eprint@arXiv#1{\href {http://arxiv.org/abs/#1} {{\tt arXiv:#1}}}
\def\mn@eprint@dblp#1{\href {http://dblp.uni-trier.de/rec/bibtex/#1.xml}
  {dblp:#1}}
\def\mn@eprint@#1:#2:#3:#4\@nil{\def\@tempa {#1}\def\@tempb {#2}\def\@tempc
  {#3}\ifx \@tempc \@empty \let \@tempc \@tempb \let \@tempb \@tempa \fi \ifx
  \@tempb \@empty \def\@tempb {arXiv}\fi \@ifundefined
  {mn@eprint@\@tempb}{\@tempb:\@tempc}{\expandafter \expandafter \csname
  mn@eprint@\@tempb\endcsname \expandafter{\@tempc}}}

\bibitem[\protect\citeauthoryear{{Begelman}, {McKee}  \& {Shields}}{{Begelman}
  et~al.}{1983}]{begelman83}
{Begelman} M.~C.,  {McKee} C.~F.,   {Shields} G.~A.,  1983, \mn@doi [\apj]
  {10.1086/161178}, \href
  {https://ui.adsabs.harvard.edu/abs/1983ApJ...271...70B} {271, 70}

\bibitem[\protect\citeauthoryear{{Bennert}, {Jungwiert}, {Komossa}, {Haas}  \&
  {Chini}}{{Bennert} et~al.}{2006}]{bennert06}
{Bennert} N.,  {Jungwiert} B.,  {Komossa} S.,  {Haas} M.,   {Chini} R.,  2006,
  \mn@doi [\aap] {10.1051/0004-6361:20065477}, \href
  {https://ui.adsabs.harvard.edu/abs/2006A&A...459...55B} {459, 55}

\bibitem[\protect\citeauthoryear{{Blandford} \& {Payne}}{{Blandford} \&
  {Payne}}{1982}]{blandford82}
{Blandford} R.~D.,  {Payne} D.~G.,  1982, \mn@doi [\mnras]
  {10.1093/mnras/199.4.883}, \href
  {https://ui.adsabs.harvard.edu/abs/1982MNRAS.199..883B} {199, 883}

\bibitem[\protect\citeauthoryear{{Boirin}, {M{\'e}ndez}, {D{\'\i}az Trigo},
  {Parmar}  \& {Kaastra}}{{Boirin} et~al.}{2005}]{boirin05}
{Boirin} L.,  {M{\'e}ndez} M.,  {D{\'\i}az Trigo} M.,  {Parmar} A.~N.,
  {Kaastra} J.~S.,  2005, \mn@doi [\aap] {10.1051/0004-6361:20041940}, \href
  {https://ui.adsabs.harvard.edu/abs/2005A&A...436..195B} {436, 195}

\bibitem[\protect\citeauthoryear{{Bright}, {Fender}, {Motta}, {Rhodes},
  {Titterington}  \& {Perrott}}{{Bright} et~al.}{2018}]{bright18}
{Bright} J.,  {Fender} R.,  {Motta} S.,  {Rhodes} L.,  {Titterington} D.,
  {Perrott} Y.,  2018, The Astronomer's Telegram, \href
  {https://ui.adsabs.harvard.edu/abs/2018ATel12184....1B} {12184, 1}

\bibitem[\protect\citeauthoryear{{Buisson}, {Altamirano}, {Remillard},
  {Arzoumanian}, {Gendreau}, {Gandhi}  \& {Vincentelli}}{{Buisson}
  et~al.}{2020a}]{buisson20atel1}
{Buisson} D. J.~K.,  {Altamirano} D.,  {Remillard} R.,  {Arzoumanian} Z.,
  {Gendreau} K.,  {Gandhi} P.,   {Vincentelli} F.,  2020a, The Astronomer's
  Telegram, \href {https://ui.adsabs.harvard.edu/abs/2020ATel13536....1B}
  {13536, 1}

\bibitem[\protect\citeauthoryear{{Buisson} et~al.,}{{Buisson}
  et~al.}{2020b}]{buisson20atel2}
{Buisson} D.~J.~K.,  et~al., 2020b, The Astronomer's Telegram, \href
  {https://ui.adsabs.harvard.edu/abs/2020ATel13563....1B} {13563, 1}

\bibitem[\protect\citeauthoryear{{Casares}, {Mu{\~n}oz-Darias}, {Mata
  S{\'a}nchez}, {Charles}, {Torres}, {Armas Padilla}, {Fender}  \&
  {Garc{\'\i}a-Rojas}}{{Casares} et~al.}{2019}]{casares19}
{Casares} J.,  {Mu{\~n}oz-Darias} T.,  {Mata S{\'a}nchez} D.,  {Charles} P.~A.,
   {Torres} M.~A.~P.,  {Armas Padilla} M.,  {Fender} R.~P.,
  {Garc{\'\i}a-Rojas} J.,  2019, \mn@doi [\mnras] {10.1093/mnras/stz1793},
  \href {https://ui.adsabs.harvard.edu/abs/2019MNRAS.488.1356C} {488, 1356}

\bibitem[\protect\citeauthoryear{{Cash}}{{Cash}}{1979}]{cash79}
{Cash} W.,  1979, \mn@doi [\apj] {10.1086/156922}, \href
  {https://ui.adsabs.harvard.edu/abs/1979ApJ...228..939C} {228, 939}

\bibitem[\protect\citeauthoryear{{Charles}, {Matthews}, {Buckley}, {Gandhi},
  {Kotze}  \& {Paice}}{{Charles} et~al.}{2019}]{charles19}
{Charles} P.,  {Matthews} J.~H.,  {Buckley} D. A.~H.,  {Gandhi} P.,  {Kotze}
  E.,   {Paice} J.,  2019, \mn@doi [\mnras] {10.1093/mnrasl/slz120}, \href
  {https://ui.adsabs.harvard.edu/abs/2019MNRAS.489L..47C} {489, L47}

\bibitem[\protect\citeauthoryear{{Costantini} et~al.,}{{Costantini}
  et~al.}{2012}]{costantini12}
{Costantini} E.,  et~al., 2012, \mn@doi [\aap] {10.1051/0004-6361/201117818},
  \href {https://ui.adsabs.harvard.edu/abs/2012A&A...539A..32C} {539, A32}

\bibitem[\protect\citeauthoryear{{Cottam}, {Sako}, {Kahn}, {Paerels}  \&
  {Liedahl}}{{Cottam} et~al.}{2001}]{cottam01}
{Cottam} J.,  {Sako} M.,  {Kahn} S.~M.,  {Paerels} F.,   {Liedahl} D.~A.,
  2001, \mn@doi [\apjl] {10.1086/323343}, \href
  {https://ui.adsabs.harvard.edu/abs/2001ApJ...557L.101C} {557, L101}

\bibitem[\protect\citeauthoryear{{D{\'\i}az Trigo} \& {Boirin}}{{D{\'\i}az
  Trigo} \& {Boirin}}{2013}]{diaztrigo13}
{D{\'\i}az Trigo} M.,  {Boirin} L.,  2013, Acta Polytechnica, \href
  {https://ui.adsabs.harvard.edu/abs/2013AcPol..53..659D} {53, 659}

\bibitem[\protect\citeauthoryear{{D{\'\i}az Trigo} \& {Boirin}}{{D{\'\i}az
  Trigo} \& {Boirin}}{2016}]{diaztrigo16}
{D{\'\i}az Trigo} M.,  {Boirin} L.,  2016, \mn@doi [Astronomische Nachrichten]
  {10.1002/asna.201612315}, \href
  {https://ui.adsabs.harvard.edu/abs/2016AN....337..368D} {337, 368}

\bibitem[\protect\citeauthoryear{{Dorodnitsyn}}{{Dorodnitsyn}}{2009}]{dorodnitsyn09}
{Dorodnitsyn} A.~V.,  2009, \mn@doi [\mnras]
  {10.1111/j.1365-2966.2008.14171.x}, \href
  {https://ui.adsabs.harvard.edu/abs/2009MNRAS.393.1433D} {393, 1433}

\bibitem[\protect\citeauthoryear{{Dorodnitsyn}}{{Dorodnitsyn}}{2010}]{dorodnitsyn10}
{Dorodnitsyn} A.~V.,  2010, \mn@doi [\mnras]
  {10.1111/j.1365-2966.2010.16728.x}, \href
  {https://ui.adsabs.harvard.edu/abs/2010MNRAS.406.1060D} {406, 1060}

\bibitem[\protect\citeauthoryear{{Fender} \& {Mu{\~n}oz-Darias}}{{Fender} \&
  {Mu{\~n}oz-Darias}}{2016}]{fender16}
{Fender} R.,  {Mu{\~n}oz-Darias} T.,  2016, {The Balance of Power: Accretion
  and Feedback in Stellar Mass Black Holes}.
p.~65, \mn@doi{10.1007/978-3-319-19416-5_3}

\bibitem[\protect\citeauthoryear{{Foreman-Mackey}, {Hogg}, {Lang}  \&
  {Goodman}}{{Foreman-Mackey} et~al.}{2013}]{foremanmackey13}
{Foreman-Mackey} D.,  {Hogg} D.~W.,  {Lang} D.,   {Goodman} J.,  2013, \mn@doi
  [\pasp] {10.1086/670067}, \href
  {https://ui.adsabs.harvard.edu/abs/2013PASP..125..306F} {125, 306}

\bibitem[\protect\citeauthoryear{{Gandhi} et~al.,}{{Gandhi}
  et~al.}{2016}]{gandhi16}
{Gandhi} P.,  et~al., 2016, \mn@doi [\mnras] {10.1093/mnras/stw571}, \href
  {https://ui.adsabs.harvard.edu/abs/2016MNRAS.459..554G} {459, 554}

\bibitem[\protect\citeauthoryear{{Gandhi}, {Rao}, {Johnson}, {Paice}  \&
  {Maccarone}}{{Gandhi} et~al.}{2019}]{gandhi19}
{Gandhi} P.,  {Rao} A.,  {Johnson} M. A.~C.,  {Paice} J.~A.,   {Maccarone}
  T.~J.,  2019, \mn@doi [\mnras] {10.1093/mnras/stz438}, \href
  {https://ui.adsabs.harvard.edu/abs/2019MNRAS.485.2642G} {485, 2642}

\bibitem[\protect\citeauthoryear{{Goodman} \& {Weare}}{{Goodman} \&
  {Weare}}{2010}]{goodman10}
{Goodman} J.,  {Weare} J.,  2010, \mn@doi [Communications in Applied
  Mathematics and Computational Science] {10.2140/camcos.2010.5.65}, \href
  {https://ui.adsabs.harvard.edu/abs/2010CAMCS...5...65G} {5, 65}

\bibitem[\protect\citeauthoryear{{Hare}, {Gandhi}, {Paice}  \&
  {Tomsick}}{{Hare} et~al.}{2019}]{hare19}
{Hare} J.,  {Gandhi} P.,  {Paice} J.~A.,   {Tomsick} J.,  2019, The
  Astronomer's Telegram, \href
  {https://ui.adsabs.harvard.edu/abs/2019ATel12512....1H} {12512, 1}

\bibitem[\protect\citeauthoryear{{Hare} et~al.,}{{Hare} et~al.}{2020}]{hare20}
{Hare} J.,  et~al., 2020, arXiv e-prints, \href
  {https://ui.adsabs.harvard.edu/abs/2020arXiv200103214H} {p. arXiv:2001.03214}

\bibitem[\protect\citeauthoryear{{Higginbottom}, {Knigge}, {Long}, {Matthews},
  {Sim}  \& {Hewitt}}{{Higginbottom} et~al.}{2018}]{higginbottom18}
{Higginbottom} N.,  {Knigge} C.,  {Long} K.~S.,  {Matthews} J.~H.,  {Sim}
  S.~A.,   {Hewitt} H.~A.,  2018, \mn@doi [\mnras] {10.1093/mnras/sty1599},
  \href {https://ui.adsabs.harvard.edu/abs/2018MNRAS.479.3651H} {479, 3651}

\bibitem[\protect\citeauthoryear{{Houck} \& {Denicola}}{{Houck} \&
  {Denicola}}{2000}]{houck00}
{Houck} J.~C.,  {Denicola} L.~A.,  2000, {ISIS: An Interactive Spectral
  Interpretation System for High Resolution X-Ray Spectroscopy}.
p.~591

\bibitem[\protect\citeauthoryear{{Icke}}{{Icke}}{1980}]{icke80}
{Icke} V.,  1980, \mn@doi [\aj] {10.1086/112678}, \href
  {https://ui.adsabs.harvard.edu/abs/1980AJ.....85..329I} {85, 329}

\bibitem[\protect\citeauthoryear{{Jansen} et~al.,}{{Jansen}
  et~al.}{2001}]{jansen01}
{Jansen} F.,  et~al., 2001, \mn@doi [\aap] {10.1051/0004-6361:20000036}, \href
  {https://ui.adsabs.harvard.edu/abs/2001A&A...365L...1J} {365, L1}

\bibitem[\protect\citeauthoryear{{Jiang}, {Fabian}, {Wang}, {Walton},
  {Garc{\'\i}a}, {Parker}, {Steiner}  \& {Tomsick}}{{Jiang}
  et~al.}{2019}]{jiang19}
{Jiang} J.,  {Fabian} A.~C.,  {Wang} J.,  {Walton} D.~J.,  {Garc{\'\i}a} J.~A.,
   {Parker} M.~L.,  {Steiner} J.~F.,   {Tomsick} J.~A.,  2019, \mn@doi [\mnras]
  {10.1093/mnras/stz095}, \href
  {https://ui.adsabs.harvard.edu/abs/2019MNRAS.484.1972J} {484, 1972}

\bibitem[\protect\citeauthoryear{{Jimenez-Garate}, {Raymond}  \&
  {Liedahl}}{{Jimenez-Garate} et~al.}{2002}]{jimenezgarate02}
{Jimenez-Garate} M.~A.,  {Raymond} J.~C.,   {Liedahl} D.~A.,  2002, \mn@doi
  [\apj] {10.1086/344364}, \href
  {https://ui.adsabs.harvard.edu/abs/2002ApJ...581.1297J} {581, 1297}

\bibitem[\protect\citeauthoryear{{Kafka} \& {Honeycutt}}{{Kafka} \&
  {Honeycutt}}{2004}]{kafka04}
{Kafka} S.,  {Honeycutt} R.~K.,  2004, \mn@doi [\aj] {10.1086/424618}, \href
  {https://ui.adsabs.harvard.edu/abs/2004AJ....128.2420K} {128, 2420}

\bibitem[\protect\citeauthoryear{{Kallman} \& {Bautista}}{{Kallman} \&
  {Bautista}}{2001}]{kallman01}
{Kallman} T.,  {Bautista} M.,  2001, \mn@doi [\apjs] {10.1086/319184}, \href
  {https://ui.adsabs.harvard.edu/abs/2001ApJS..133..221K} {133, 221}

\bibitem[\protect\citeauthoryear{{Kallman}, {Angelini}, {Boroson}  \&
  {Cottam}}{{Kallman} et~al.}{2003}]{kallman03}
{Kallman} T.~R.,  {Angelini} L.,  {Boroson} B.,   {Cottam} J.,  2003, \mn@doi
  [\apj] {10.1086/345475}, \href
  {https://ui.adsabs.harvard.edu/abs/2003ApJ...583..861K} {583, 861}

\bibitem[\protect\citeauthoryear{{Kallman}, {Bautista}, {Goriely}, {Mendoza},
  {Miller}, {Palmeri}, {Quinet}  \& {Raymond}}{{Kallman}
  et~al.}{2009}]{kallman09}
{Kallman} T.~R.,  {Bautista} M.~A.,  {Goriely} S.,  {Mendoza} C.,  {Miller}
  J.~M.,  {Palmeri} P.,  {Quinet} P.,   {Raymond} J.,  2009, \mn@doi [\apj]
  {10.1088/0004-637X/701/2/865}, \href
  {https://ui.adsabs.harvard.edu/abs/2009ApJ...701..865K} {701, 865}

\bibitem[\protect\citeauthoryear{{King} et~al.,}{{King} et~al.}{2012}]{king12}
{King} A.~L.,  et~al., 2012, \mn@doi [\apjl] {10.1088/2041-8205/746/2/L20},
  \href {https://ui.adsabs.harvard.edu/abs/2012ApJ...746L..20K} {746, L20}

\bibitem[\protect\citeauthoryear{{King}, {Miller}, {Raymond}, {Reynolds}  \&
  {Morningstar}}{{King} et~al.}{2015}]{king15}
{King} A.~L.,  {Miller} J.~M.,  {Raymond} J.,  {Reynolds} M.~T.,
  {Morningstar} W.,  2015, \mn@doi [\apjl] {10.1088/2041-8205/813/2/L37}, \href
  {https://ui.adsabs.harvard.edu/abs/2015ApJ...813L..37K} {813, L37}

\bibitem[\protect\citeauthoryear{{Krimm} et~al.,}{{Krimm}
  et~al.}{2018}]{krimm18}
{Krimm} H.~A.,  et~al., 2018, The Astronomer's Telegram, \href
  {https://ui.adsabs.harvard.edu/abs/2018ATel12151....1K} {12151, 1}

\bibitem[\protect\citeauthoryear{{Mauche}, {Liedahl}  \& {Fournier}}{{Mauche}
  et~al.}{2003}]{mauche03}
{Mauche} C.~W.,  {Liedahl} D.~A.,   {Fournier} K.~B.,  2003, \mn@doi [\apjl]
  {10.1086/375684}, \href
  {https://ui.adsabs.harvard.edu/abs/2003ApJ...588L.101M} {588, L101}

\bibitem[\protect\citeauthoryear{{Miller} \& {Stone}}{{Miller} \&
  {Stone}}{2000}]{miller00}
{Miller} K.~A.,  {Stone} J.~M.,  2000, \mn@doi [\apj] {10.1086/308736}, \href
  {https://ui.adsabs.harvard.edu/abs/2000ApJ...534..398M} {534, 398}

\bibitem[\protect\citeauthoryear{{Miller}, {Raymond}, {Fabian}, {Steeghs},
  {Homan}, {Reynolds}, {van der Klis}  \& {Wijnands}}{{Miller}
  et~al.}{2006a}]{miller06nat}
{Miller} J.~M.,  {Raymond} J.,  {Fabian} A.,  {Steeghs} D.,  {Homan} J.,
  {Reynolds} C.,  {van der Klis} M.,   {Wijnands} R.,  2006a, \mn@doi [\nat]
  {10.1038/nature04912}, \href
  {https://ui.adsabs.harvard.edu/abs/2006Natur.441..953M} {441, 953}

\bibitem[\protect\citeauthoryear{{Miller} et~al.,}{{Miller}
  et~al.}{2006b}]{miller061743}
{Miller} J.~M.,  et~al., 2006b, \mn@doi [\apj] {10.1086/504673}, \href
  {https://ui.adsabs.harvard.edu/abs/2006ApJ...646..394M} {646, 394}

\bibitem[\protect\citeauthoryear{{Miller}, {Raymond}, {Reynolds}, {Fabian},
  {Kallman}  \& {Homan}}{{Miller} et~al.}{2008}]{miller08}
{Miller} J.~M.,  {Raymond} J.,  {Reynolds} C.~S.,  {Fabian} A.~C.,  {Kallman}
  T.~R.,   {Homan} J.,  2008, \mn@doi [\apj] {10.1086/588521}, \href
  {https://ui.adsabs.harvard.edu/abs/2008ApJ...680.1359M} {680, 1359}

\bibitem[\protect\citeauthoryear{{Miller} et~al.,}{{Miller}
  et~al.}{2014}]{miller14}
{Miller} J.~M.,  et~al., 2014, \mn@doi [\apj] {10.1088/0004-637X/788/1/53},
  \href {https://ui.adsabs.harvard.edu/abs/2014ApJ...788...53M} {788, 53}

\bibitem[\protect\citeauthoryear{{Miller}, {Fabian}, {Kaastra}, {Kallman},
  {King}, {Proga}, {Raymond}  \& {Reynolds}}{{Miller} et~al.}{2015}]{miller15}
{Miller} J.~M.,  {Fabian} A.~C.,  {Kaastra} J.,  {Kallman} T.,  {King} A.~L.,
  {Proga} D.,  {Raymond} J.,   {Reynolds} C.~S.,  2015, \mn@doi [\apj]
  {10.1088/0004-637X/814/2/87}, \href
  {https://ui.adsabs.harvard.edu/abs/2015ApJ...814...87M} {814, 87}

\bibitem[\protect\citeauthoryear{{Miller} et~al.,}{{Miller}
  et~al.}{2016}]{miller16}
{Miller} J.~M.,  et~al., 2016, \mn@doi [\apjl] {10.3847/2041-8205/821/1/L9},
  \href {https://ui.adsabs.harvard.edu/abs/2016ApJ...821L...9M} {821, L9}

\bibitem[\protect\citeauthoryear{{Motta}, {Kajava},
  {S{\'a}nchez-Fern{\'a}ndez}, {Giustini}  \& {Kuulkers}}{{Motta}
  et~al.}{2017}]{motta17abs}
{Motta} S.~E.,  {Kajava} J.~J.~E.,  {S{\'a}nchez-Fern{\'a}ndez} C.,  {Giustini}
  M.,   {Kuulkers} E.,  2017, \mn@doi [\mnras] {10.1093/mnras/stx466}, \href
  {https://ui.adsabs.harvard.edu/abs/2017MNRAS.468..981M} {468, 981}

\bibitem[\protect\citeauthoryear{{Mu{\~n}oz-Darias} et~al.,}{{Mu{\~n}oz-Darias}
  et~al.}{2016}]{munozdarias16}
{Mu{\~n}oz-Darias} T.,  et~al., 2016, \mn@doi [\nat] {10.1038/nature17446},
  \href {https://ui.adsabs.harvard.edu/abs/2016Natur.534...75M} {534, 75}

\bibitem[\protect\citeauthoryear{{Mu{\~n}oz-Darias}, {Torres}  \&
  {Garcia}}{{Mu{\~n}oz-Darias} et~al.}{2018}]{munozdarias18}
{Mu{\~n}oz-Darias} T.,  {Torres} M. A.~P.,   {Garcia} M.~R.,  2018, \mn@doi
  [\mnras] {10.1093/mnras/sty1711}, \href
  {https://ui.adsabs.harvard.edu/abs/2018MNRAS.479.3987M} {479, 3987}

\bibitem[\protect\citeauthoryear{{Mu{\~n}oz-Darias} et~al.,}{{Mu{\~n}oz-Darias}
  et~al.}{2020}]{munoz20}
{Mu{\~n}oz-Darias} T.,  et~al., 2020, arXiv e-prints, \href
  {https://ui.adsabs.harvard.edu/abs/2020arXiv200312073M} {p. arXiv:2003.12073}

\bibitem[\protect\citeauthoryear{{Munoz-Darias}, {Jimenez-Ibarra}, {Armas
  Padilla}, {Casares}, {Cuneo}, {Panizo-Espinar}, {Sanchez-Sierras}  \&
  {Torres}}{{Munoz-Darias} et~al.}{2019}]{munoz19}
{Munoz-Darias} T.,  {Jimenez-Ibarra} F.,  {Armas Padilla} M.,  {Casares} J.,
  {Cuneo} V.,  {Panizo-Espinar} G.,  {Sanchez-Sierras} J.,   {Torres} M.~A.~P.,
   2019, The Astronomer's Telegram, \href
  {https://ui.adsabs.harvard.edu/abs/2019ATel12881....1M} {12881, 1}

\bibitem[\protect\citeauthoryear{{Neilsen}, {Remillard}  \& {Lee}}{{Neilsen}
  et~al.}{2011}]{neilsen11}
{Neilsen} J.,  {Remillard} R.~A.,   {Lee} J.~C.,  2011, \mn@doi [\apj]
  {10.1088/0004-637X/737/2/69}, \href
  {https://ui.adsabs.harvard.edu/abs/2011ApJ...737...69N} {737, 69}

\bibitem[\protect\citeauthoryear{{Paice}, {Gandhi}, {Dhillon}, {Marsh}, {Green}
   \& {Breedt}}{{Paice} et~al.}{2018}]{paice18}
{Paice} J.~A.,  {Gandhi} P.,  {Dhillon} V.~S.,  {Marsh} T.~R.,  {Green} M.,
  {Breedt} E.,  2018, The Astronomer's Telegram, \href
  {https://ui.adsabs.harvard.edu/abs/2018ATel12197....1P} {12197, 1}

\bibitem[\protect\citeauthoryear{{Ponti}, {Fender}, {Begelman}, {Dunn},
  {Neilsen}  \& {Coriat}}{{Ponti} et~al.}{2012}]{ponti12}
{Ponti} G.,  {Fender} R.~P.,  {Begelman} M.~C.,  {Dunn} R.~J.~H.,  {Neilsen}
  J.,   {Coriat} M.,  2012, \mn@doi [\mnras]
  {10.1111/j.1745-3933.2012.01224.x}, \href
  {https://ui.adsabs.harvard.edu/abs/2012MNRAS.422L..11P} {422, L11}

\bibitem[\protect\citeauthoryear{{Porquet}, {Mewe}, {Dubau}, {Raassen}  \&
  {Kaastra}}{{Porquet} et~al.}{2001}]{porquet01}
{Porquet} D.,  {Mewe} R.,  {Dubau} J.,  {Raassen} A.~J.~J.,   {Kaastra} J.~S.,
  2001, \mn@doi [\aap] {10.1051/0004-6361:20010959}, \href
  {https://ui.adsabs.harvard.edu/abs/2001A&A...376.1113P} {376, 1113}

\bibitem[\protect\citeauthoryear{{Prinja}, {Ringwald}, {Wade}  \&
  {Knigge}}{{Prinja} et~al.}{2000}]{prinja00}
{Prinja} R.~K.,  {Ringwald} F.~A.,  {Wade} R.~A.,   {Knigge} C.,  2000, \mn@doi
  [\mnras] {10.1046/j.1365-8711.2000.03111.x}, \href
  {https://ui.adsabs.harvard.edu/abs/2000MNRAS.312..316P} {312, 316}

\bibitem[\protect\citeauthoryear{{Proga} \& {Kallman}}{{Proga} \&
  {Kallman}}{2002}]{proga02}
{Proga} D.,  {Kallman} T.~R.,  2002, \mn@doi [\apj] {10.1086/324534}, \href
  {https://ui.adsabs.harvard.edu/abs/2002ApJ...565..455P} {565, 455}

\bibitem[\protect\citeauthoryear{{Protassov}, {van Dyk}, {Connors}, {Kashyap}
  \& {Siemiginowska}}{{Protassov} et~al.}{2002}]{protassov02}
{Protassov} R.,  {van Dyk} D.~A.,  {Connors} A.,  {Kashyap} V.~L.,
  {Siemiginowska} A.,  2002, \mn@doi [\apj] {10.1086/339856}, \href
  {http://adsabs.harvard.edu/abs/2002ApJ...571..545P} {571, 545}

\bibitem[\protect\citeauthoryear{{Psaradaki}, {Costantini}, {Mehdipour}  \&
  {D{\'\i}az Trigo}}{{Psaradaki} et~al.}{2018}]{psaradaki18}
{Psaradaki} I.,  {Costantini} E.,  {Mehdipour} M.,   {D{\'\i}az Trigo} M.,
  2018, \mn@doi [\aap] {10.1051/0004-6361/201834000}, \href
  {https://ui.adsabs.harvard.edu/abs/2018A&A...620A.129P} {620, A129}

\bibitem[\protect\citeauthoryear{{Puebla}, {Diaz}, {Hillier}  \&
  {Hubeny}}{{Puebla} et~al.}{2011}]{puebla11}
{Puebla} R.~E.,  {Diaz} M.~P.,  {Hillier} D.~J.,   {Hubeny} I.,  2011, \mn@doi
  [\apj] {10.1088/0004-637X/736/1/17}, \href
  {https://ui.adsabs.harvard.edu/abs/2011ApJ...736...17P} {736, 17}

\bibitem[\protect\citeauthoryear{{Rajwade} et~al.,}{{Rajwade}
  et~al.}{2019}]{rajwade19}
{Rajwade} K.~M.,  et~al., 2019, The Astronomer's Telegram, \href
  {https://ui.adsabs.harvard.edu/abs/2019ATel12499....1R} {12499, 1}

\bibitem[\protect\citeauthoryear{{Revnivtsev}, {Gilfanov}, {Churazov}  \&
  {Sunyaev}}{{Revnivtsev} et~al.}{2002}]{revnivtsev02}
{Revnivtsev} M.,  {Gilfanov} M.,  {Churazov} E.,   {Sunyaev} R.,  2002, \mn@doi
  [\aap] {10.1051/0004-6361:20020865}, \href
  {https://ui.adsabs.harvard.edu/abs/2002A&A...391.1013R} {391, 1013}

\bibitem[\protect\citeauthoryear{{Reynolds}, {Miller}, {Ludlam}  \&
  {Tetarenko}}{{Reynolds} et~al.}{2018}]{reynolds18}
{Reynolds} M.~T.,  {Miller} J.~M.,  {Ludlam} R.~M.,   {Tetarenko} B.~E.,  2018,
  The Astronomer's Telegram, \href
  {https://ui.adsabs.harvard.edu/abs/2018ATel12220....1R} {12220, 1}

\bibitem[\protect\citeauthoryear{{Schulz} \& {Brandt}}{{Schulz} \&
  {Brandt}}{2002}]{schulz02}
{Schulz} N.~S.,  {Brandt} W.~N.,  2002, \mn@doi [\apj] {10.1086/340369}, \href
  {https://ui.adsabs.harvard.edu/abs/2002ApJ...572..971S} {572, 971}

\bibitem[\protect\citeauthoryear{{Svensson} \& {Zdziarski}}{{Svensson} \&
  {Zdziarski}}{1994}]{svensson94}
{Svensson} R.,  {Zdziarski} A.~A.,  1994, \mn@doi [\apj] {10.1086/174934},
  \href {https://ui.adsabs.harvard.edu/abs/1994ApJ...436..599S} {436, 599}

\bibitem[\protect\citeauthoryear{{Tomsick} et~al.,}{{Tomsick}
  et~al.}{2018}]{tomsick18}
{Tomsick} J.~A.,  et~al., 2018, \mn@doi [\apj] {10.3847/1538-4357/aaaab1},
  \href {https://ui.adsabs.harvard.edu/abs/2018ApJ...855....3T} {855, 3}

\bibitem[\protect\citeauthoryear{{Tremaine}, {Shen}, {Liu}  \&
  {Loeb}}{{Tremaine} et~al.}{2014}]{tremaine14}
{Tremaine} S.,  {Shen} Y.,  {Liu} X.,   {Loeb} A.,  2014, \mn@doi [\apj]
  {10.1088/0004-637X/794/1/49}, \href
  {https://ui.adsabs.harvard.edu/abs/2014ApJ...794...49T} {794, 49}

\bibitem[\protect\citeauthoryear{{Ueda}, {Inoue}, {Tanaka}, {Ebisawa},
  {Nagase}, {Kotani}  \& {Gehrels}}{{Ueda} et~al.}{1998}]{ueda98}
{Ueda} Y.,  {Inoue} H.,  {Tanaka} Y.,  {Ebisawa} K.,  {Nagase} F.,  {Kotani}
  T.,   {Gehrels} N.,  1998, \mn@doi [\apj] {10.1086/305063}, \href
  {https://ui.adsabs.harvard.edu/abs/1998ApJ...492..782U} {492, 782}

\bibitem[\protect\citeauthoryear{{Vasilopoulos}, {Bailyn}  \&
  {Milburn}}{{Vasilopoulos} et~al.}{2018}]{vasilopoulos18}
{Vasilopoulos} G.,  {Bailyn} C.,   {Milburn} J.,  2018, The Astronomer's
  Telegram, \href {https://ui.adsabs.harvard.edu/abs/2018ATel12164....1V}
  {12164, 1}

\bibitem[\protect\citeauthoryear{{Walton} et~al.,}{{Walton}
  et~al.}{2017}]{walton17}
{Walton} D.~J.,  et~al., 2017, \mn@doi [\apj] {10.3847/1538-4357/aa67e8}, \href
  {https://ui.adsabs.harvard.edu/abs/2017ApJ...839..110W} {839, 110}

\bibitem[\protect\citeauthoryear{{White} \& {Holt}}{{White} \&
  {Holt}}{1982}]{white82}
{White} N.~E.,  {Holt} S.~S.,  1982, \mn@doi [\apj] {10.1086/159991}, \href
  {https://ui.adsabs.harvard.edu/abs/1982ApJ...257..318W} {257, 318}

\bibitem[\protect\citeauthoryear{{Wijnands} \& {van der Klis}}{{Wijnands} \&
  {van der Klis}}{2000}]{wijnands00}
{Wijnands} R.,  {van der Klis} M.,  2000, \mn@doi [\apjl] {10.1086/312439},
  \href {https://ui.adsabs.harvard.edu/abs/2000ApJ...528L..93W} {528, L93}

\bibitem[\protect\citeauthoryear{{Woods}, {Klein}, {Castor}, {McKee}  \&
  {Bell}}{{Woods} et~al.}{1996}]{woods96}
{Woods} D.~T.,  {Klein} R.~I.,  {Castor} J.~I.,  {McKee} C.~F.,   {Bell} J.~B.,
   1996, \mn@doi [\apj] {10.1086/177101}, \href
  {https://ui.adsabs.harvard.edu/abs/1996ApJ...461..767W} {461, 767}

\bibitem[\protect\citeauthoryear{{van den Eijnden} et~al.,}{{van den Eijnden}
  et~al.}{2020}]{vandeneijnden20}
{van den Eijnden} J.,  et~al., 2020, \mn@doi [\mnras] {10.1093/mnras/staa1704},
  \href {https://ui.adsabs.harvard.edu/abs/2020MNRAS.496.4127V} {496, 4127}

\makeatother
\end{thebibliography}

\bsp	
\label{lastpage}

\end{document}